\documentclass[aps,pre,twocolumn,showpacs,superscriptaddress,groupedaddress]{revtex4}
\usepackage{graphicx}
\usepackage{bm}        
\usepackage{amssymb}   
\begin{document}

\title{Electric-field-driven polymer entry into asymmetric nanoscale channels }

\author{Narges Nikoofard}
\affiliation{Department of Physics, Institute for Advanced Studies in Basic Sciences (IASBS), Zanjan 45137-66731, Iran}
\author{Hossein Fazli}
\email{fazli@iasbs.ac.ir}
\affiliation{Department of Physics, Institute for Advanced Studies in Basic Sciences (IASBS), Zanjan 45137-66731, Iran}
\affiliation{Department of Biological Sciences, Institute for Advanced Studies in Basic Sciences (IASBS), Zanjan 45137-66731, Iran}

\date{\today}

\begin{abstract}
The electric-field-driven entry process of flexible charged polymers such as single stranded DNA (ssDNA) into asymmetric nanoscale channels such as $\alpha$-hemolysin protein channel is studied theoretically and using molecular dynamics simulations. Dependence of the height of the free-energy barrier on the polymer length, the strength of the applied electric field and the channel entrance geometry is investigated. It is shown that the squeezing effect of the driving field on the polymer and the lateral confinement of the polymer before its entry to the channel crucially affect the barrier height and its dependence on the system parameters. The attempt frequency of the polymer for passing the channel is also discussed. Our theoretical and simulation results support each other and describe related data sets of polymer translocation experiments through the $\alpha$-hemolysin protein channel reasonably well.
\end{abstract}

\maketitle

\section{Introduction}
Polymer translocation -- passage of a polymer through a pore in a membrane -- is a very ubiquitous and vital process in biological environments. In the past decade it has also been used as a tool for single-molecule studies \cite{single-molecule}. Better understanding of related biological processes, DNA sequencing  \cite{sequencing}, examining the existing theories on the statics and dynamics of confined polymers \cite{partitioning1,confined}, direct study of nucleic acid-protein interactions \cite{protein} and nucleic acids secondary structure \cite{2structure} are examples of present and potential applications. As a polymer passes through a pore, the number of its possible conformations and hence its entropy decreases. To overcome this entropic barrier, a voltage difference is often applied in the case of charged polymers. During the polymer passage, the ion current through the channel is blocked showing that the polymer is inside the channel \cite{voltage}. Time length of polymer passage through the pore \cite{sung-muthu,passage-time}, ion current during the polymer passage \cite{current,electrostatics} and the time interval between consecutive passages (entry time) \cite{electrostatics,meller2010,fluorescence,golovchenko,muthukumar2010,grosberg,ambjornsson,
henrickson,fazli,muthukumar2009,attempt-time,muthukumar2007,bayley,low-salt,salt-gradient,
meller-branton,pelta-auvray,trans-cis,chen} are three measurable quantities in polymer translocation experiments. These quantities contain information about the polymer and its interaction with the channel \cite{meller-review}. Despite the two first quantities, the polymer entry time into the channel is less investigated.

The entry process of the polymer into the channel may be diffusion- or barrier-limited. It has been observed in the experiments that the entry time as a function of the inverse of the applied voltage is linear (exponential) in diffusion-limited (barrier-limited) regime \cite{meller2010}. Fluorescence microscopy has revealed that in the diffusion-limited regime, there is a region of very low polymer concentration around the channel entrance. In this region, the electric field drift dominates over the diffusion and the polymer is carried to the pore in a very short time. This is continued with immediate passage of the polymer through the pore \cite{fluorescence}. In this regime, diffusion-convection (Smoluchowski) equation can explain the behavior of the polymer outside the channel and its flux through the pore very well \cite{golovchenko,muthukumar2010,grosberg,ambjornsson}.

In the barrier-limited regime, the entry process of the polymer into the channel is very time consuming. The free-energy barrier against the polymer entry into the channel \cite{fazli,grosberg,muthukumar2009} and the polymer attempt frequency for passing this barrier \cite{grosberg,attempt-time} are believed as two main quantities determining the entry time \cite{henrickson}. It has been shown that in this regime the entry time depends on the polymer length in contrast to the diffusion-limited regime \cite{meller2010}.

Positional distribution of charges on the channel surface, electro-osmotic flow of counterions, interaction of the charges on the polymer with the channel medium which is of low dielectric constant and compression of the cloud of the polymer counterions in the course of polymer passage through the channel have been mentioned as other possible relevant factors \cite{electrostatics,muthukumar2007,bayley}.

Enhancement of the frequency of polymer entry is of great interest in polymer translocation experiments. It provides the possibility of polymer translocation studies in lower salt concentrations \cite{low-salt}, lower applied voltages \cite{golovchenko} and smaller sample amounts \cite{meller2010}. For this purpose, numerous procedures such as applying a salt gradient across the channel \cite{salt-gradient,meller2010}, charge manipulation of the protein channels by site-directed mutagenesis \cite{bayley} and layer deposition on solid state channels \cite{fluorescence} have been used.

In this paper, we investigate the polymer entry time in the barrier-limited regime for two different geometries of the channel entrance. A simple scaling theory as well as the results of coarse-grained molecular dynamics (MD) simulations are presented. We show that the electric field compresses the polymer on the wall before its entry to the channel. We calculate the free-energy of the polymer in the compressed state and then find the dependence of the height of the free-energy barrier on the polymer length and strength of the electric field for the two geometries of the channel entrance. Results of the theory and the simulation are in good agreement and their correspondence with experimental results is investigated and discussed. We also study the polymer attempt frequency for crossing the barrier. Our studies show that the squeezing effect of the electric field and the lateral confinement of the polymer prior to its entry to the channel are very important factors in the height of the free-energy barrier and the dynamics of the polymer translocation. These effects were not taken into account in the previous works. We also study the polymer entry into an asymmetric channel of $\alpha$-hemolysin dimensions and discuss important factors in the entry time and compare our results with related experiments. Here, in addition to extension of Ref. \cite{fazli}, new results on the polymer entry into an asymmetric channel of the same dimensions as the $\alpha$-hemolysin channel and the polymer attempt time for passing the channel are presented.

\begin{figure}
\includegraphics[scale=0.35]{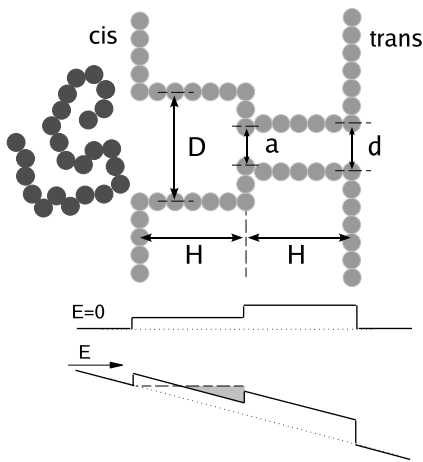}
\caption{\label{fig:epsart} Schematic of an asymmetric channel
consisting of two coaxial cylinders of different diameters and a polymer which is
released from equilibrium in front of the channel (upper panel). Schematic of the free-energy landscape of the polymer when it enters from the cis side of the channel without and with applied driving electric field (lower panel).} \label{fig1}
\end{figure}
The rest of the paper is organized as follows. The model and the method of the simulation of polymer entry into an asymmetric channel is discussed in Sec. \ref{model}. Simulation results are also presented in this section. A simple scaling theory for statics of a polymer before its entry to a channel and the height of free-energy barrier against its entry is developed in Sec. \ref{theory} and then compared with simulation results in Sec. \ref{simulation}. Simulation results for the polymer attempt time prior to the channel entrance is presented and discussed in Sec. \ref{simulation}. Section \ref{experiment} is devoted to the comparison of our results with related experimental data sets published in the literature. Finally, a brief review of the paper and some final notes are given in Sec. \ref{summary}.

\section{The Model and The Simulation Method}   \label{model}
The $\alpha$-hemolysin protein channel has a completely known crystallographic structure. It is very stable under the effect of the applied voltages and different chemicals. Thanks to such characteristics, this protein channel has been used in many polymer translocation experiments.
In these experiments, it has been observed that the polymer entry time from the trans side of the channel is considerably longer than that of the cis side. For this difference, three main reasons have been suggested: difference between channel vestibules size and so different entropic barriers against the polymer entry from the two sides, asymmetric distribution of charges on the channel which leads to polymer attraction or repulsion from the bulk \cite{henrickson} and different projections of the channel relative to the membrane and so entropically easier approach of the polymer to the cis side of the channel \cite{ambjornsson}.

To investigate the effect of the difference between the size of channel vestibules, we perform MD simulations of electric-field-driven polymer entry into an asymmetric channel of a simple geometry as shown in Fig. \ref{fig1}. In our simulations, surfaces of the flat wall containing the channel and the channel itself are constructed by spherical fixed particles of diameter $\sigma$. The walls are constructed by particles arranged on concentric circles of different radii and the two cylinders forming the channel (which are of diameters $D$ and $d$) are made by particles arranged on consecutive circles. 3D shape of the channel in our simulations can be imagined by rotating the schematic shown in Fig. \ref{fig1} around its symmetry axis.

The polymer is modeled as a bead-spring chain. The beads are of diameter $\sigma$ and excluded volume interactions of the monomers with each other and with the walls and the channel are modeled by the shifted and truncated Lennard-Jones potential,
\begin{equation}
 U_{LJ}(r) = \left\lbrace
  \begin{array}{l l}
    4\varepsilon\
    \{(\frac{\sigma}{r})^{12}-(\frac{\sigma}{r})^{6}+\frac{1}{4}\} & \text{if $r<r_{c}$},\\
    0 & \text{if $r \geq r_{c}$}.
  \end{array}
\right.
\label{lj}
\end{equation}
$r$ is the distance between the particles. $\varepsilon$ and $\sigma$ represent the strength and the length scale of the interaction, respectively and the range of interaction is $r_c = 2^{\frac{1}{6}}\sigma$. The monomers are connected by the finite extensible nonlinear elastic (FENE) potential,
\begin{equation}
U_{bond}(r)=\left\lbrace
     \begin{array}{l l}
       - \frac{1}{2}kR_{0}^{2}\ln(1-(\frac{r}{R_{0}})^{2})& \text{if $r<R_{0}$},\\
       0 & \text{if $r \geq R_{0}$}.
  \end{array}
\right.
\end{equation}
We set the spring constant and the maximum bond length as $k=70.0\frac{\varepsilon}{\sigma^2}$ and $R_0=1.5\sigma$.

The equations of motion are integrated with the velocity Verlet algorithm with the step size $0.01\tau_{MD}$, where $\tau_{MD}=\sigma\sqrt{\frac{m}{\varepsilon}}$ is the MD time scale. The system is kept at the constant temperature $T=1.0\frac{\varepsilon}{k_B}$ using the Langevin thermostat with the damping constant $1.0\tau_{MD}^{-1}$ \cite{thermostat}. So, hydrodynamic interactions are not included in our model. It has been shown that consideration of the hydrodynamic interactions in simulation do not change the polymer equilibrium properties considerably. As the entry time is very long in the barrier-limited regime, this system is in a quasi-equilibrium state.

The polymer translocation experiments are often done in $1M$ concentration of monovalent salt \cite{voltage}. The Debye screening length corresponding to such a salt concentration is $\simeq 0.3 nm$, which is smaller than the diameter of ssDNA, $1 nm$. Accordingly, we assume here that there is no electrostatic interaction between the monomers and they only interact with the applied electric field. Also no counterions or coions are considered in our model.

The electric field effect on the monomers is modeled by applying the force $\vec{F}=F\hat{z}$ to all of the monomers. This force is uniform in space and it is the same for all the monomers. All the simulations are performed with ESPResSo \cite{espresso}.

\begin{figure}
\includegraphics[scale=0.5]{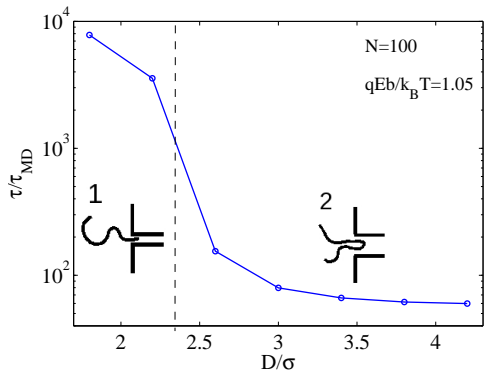}
\caption{ (Color online) Polymer entry time into a cylindrical channel in a wall versus the channel diameter. In the first two points of the plot, polymer enters from one end. In the other points, entry in folded state is also possible.}
\label{fig2}
\end{figure}

To measure the polymer entry time into a channel in a wall, we fix the middle monomer of a polymer consisting of 25 monomer at a close distance from the wall and leave the polymer to equilibrate without any applied electric field. Then we release the fixed monomer and turn on the electric field at the same time, $t=0$. The polymer entry time, $\tau$, is defined as the time at which one of the monomers completely passes the channel. In these simulations, a cylindrical constraint of diameter $D_c=14.0\sigma$ (coaxial with the two cylinders forming the channel) is set around the channel to keep the polymer from diffusing away. The diameter of this constraint is larger than the radius of gyration of the polymer to avoid any confinement effects. Excluded volume interaction of monomers with this constraint is also given by Eq. \ref{lj}. 

Presence of this constraint is justified by the fact that we are interested in the barrier-limited regime. In this regime, the time it takes the polymer to diffuse from the bulk to the channel is much shorter than the time it takes the polymer to enter into the channel. In fact, we assume that keeping a fluctuating polymer close to the pore by a constraint is equivalent to continuous diffusion of polymers of different configurations from the bulk to the channel vicinity. As a result, to compare entry times obtained from our simulations to experimental ones, they should be multiplied by the ratio of diffusion time of the polymers from the bulk to the characteristic time of the polymer reconfiguration.
One can obtain an estimation for this conversion factor. The characteristic time of polymer reconfiguration in our simulations is given with the Rouse time $\tau_R=\frac{N^{2.2}}{6\pi^2}\tau_0$ \cite{rouse}, in which $\tau_0$ is equal to the MD time scale. For a polymer with $N=25$, this time is $\tau_R\simeq\tau_{MD}$. For a ssDNA with several tens of bases, the diffusion coefficient of the polymer is $D=10^{-7}\frac{cm^2}{s}$. Considering the typical polymer concentrations in the experiments ($\sim \mu M$), the diffusion time of the polymers to the channel is around $1ms$. So the factor for converting the simulation times to experimental ones becomes $F=\frac{1ms}{\tau_{MD}}$.

To obtain the entry time for each given set of parameters, averages over 100-150 realizations are calculated. Each realization for narrower channels typically takes several hours with our machines.

\begin{figure}
\includegraphics[scale=0.5]{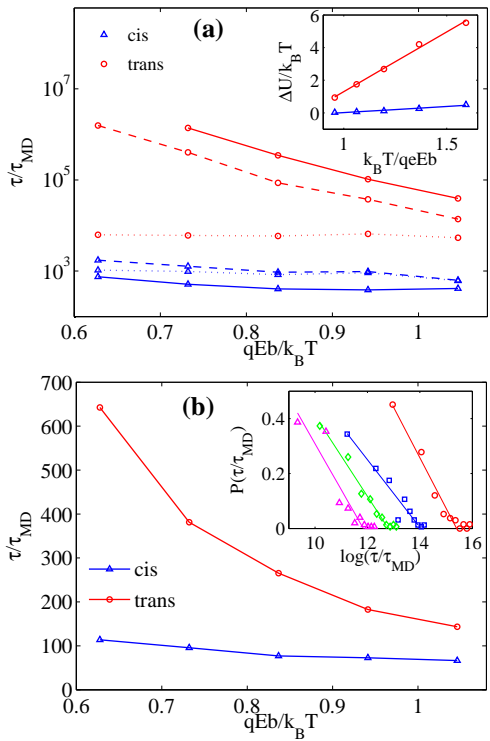}
\caption{ (a) (Color online) Entry time, $\tau$, the time it takes the polymer ends to find the channel, $\tau_0$, and $\tau_0\exp(\frac{\Delta U}{k_BT})$ are shown respectively with the solid, dotted and dashed lines. Inset: free-energy barrier versus the inverse of the electric field. (b) Entry times into a channel of the same dimensions as $\alpha$-hemplysin. Order of magnitude of the entry times and ratio entry times from cis and trans sides are in agreement with experiment \cite{henrickson}. Inset: Semi-log plot of the histogram of the entry time, $\tau$, corresponding to data points of panel (a) for the trans side. Solid lines are guides to eye showing that semi-log plot of the histograms are almost linear.}
\label{fig3}
\end{figure}

In Fig. \ref{fig2}, MD simulation result for the entry time of a polymer into a simple cylindrical channel in a wall versus the channel diameter is shown. Region 1 in this figure corresponds to the situation in which the polymer can enter the channel only by one of its ends. Region 2 however, corresponds to the other situation in which entry of the polymer in folded state is also possible. Note that there is a two orders of magnitude difference between polymer entry times in regions 1 and 2 of this figure. Entry in the folded state is impossible when the diameter of the channel is so small that two monomers can not pass the channel simultaneously or the energy barrier for entry in this state is extremely high. In rest of the paper we only consider channels that their narrow part corresponds to region 1 of Fig. \ref{fig2}.

The polymer entry time, $\tau$, from cis and trans sides of an asymmetric channel are shown in Fig. \ref{fig3}(a). Channel parameters are taken as $d=a=1.4\sigma$, $D=8\sigma$ and $H=6\sigma$. In the trans side, the polymer feels no lateral confinement before its entry to the channel. In the cis side however, the polymer is confined in a wider cylinder before its entry to the channel narrow part. Using this geometry of the channel we can probe how the presence of a wider region before the narrow part of the channel affects the polymer entry time. As it can be seen in Fig. \ref{fig3}(a), existence of the wide part in the cis side reduces the polymer entry time relative to the trans side dramatically.

Simulation results for a channel of dimensions very close to those of the $\alpha$-hemolysin channel, $d=2\sigma$, $D=3\sigma$, $H=5\sigma$ and $a=1.5\sigma$, are shown in Fig. \ref{fig3}(b). Order of magnitude of the entry times and the ratio of entry times from the cis and the trans sides are very close to the experiment \cite{henrickson}. It should be noted that a polymer containing 25 monomers in our simulations is equivalent to a 75-nucleotide ssDNA in the experiment. For more details see Sec. \ref{exp-entry}.

Our simulations show that the time-limiting stage in the process of polymer entry from the cis side is the entry from the wide part of the channel (the part of diameter $D$ in Fig. \ref{fig1}) into the narrow part. As it can be seen in Fig. \ref{fig2}, the time needed for entry in the wide part can be ignored. Entry time from both cis and trans sides can be considered as the time it takes one of the polymer ends to come in front of the channel narrow part by thermal fluctuations, $\tau_0$, multiplied by inverse of the probability to overcome the free-energy barrier of height $\Delta U$. This is the well known Van’t Hoff-Arrhenius law \cite{henrickson},

\begin{equation} \label{arrhenius}
\tau = \tau_0 \exp\left(\frac{\Delta U}{k_BT} \right).
\end{equation}

In situations that the polymer can enter into the channel in folded state, $\tau_0$ is the characteristic time of polymer fluctuations in front of the channel \cite{park}. In our case that only entry of the polymer by one end is possible, $\tau_0$ is the characteristic time it takes one of the polymer ends to reach the channel entrance.

In our simulations, $\tau_0$ and $\Delta U$ are measured separately. To calculate $\tau_0$, we block the channel narrow part and let the polymer to equilibrate under the electric field in front of the channel. Then we measure the average of the time interval between consecutive reachings of the polymer ends to the channel. When one of the end monomers of the polymer falls in a range of distance from the entrance of the channel narrow part, one time for reaching of the polymer ends to the channel is counted. For the same end of the polymer another reaching is counted only if it leaves the mentioned range and returns back again.
The amount of the equilibration time depends on the polymer length and the electric field strength. It is longer for longer polymers and weaker electric fields. We take the equilibration time when the radius of gyration of the polymer gets constant and shows only small fluctuations around the equilibrium value. For every value of $\tau_0$, 100-150 simulations are performed.

To measure $\Delta U$, we fix one of the end monomers of the polymer at the entry of the channel and let the rest of the monomers to equilibrate under the electric field. Then we release the end monomer and probe whether it enters the channel narrow part or returns back. The logarithm of the probability of successful entry (ratio of the number of successful entries to the total number of simulations) equals to $-\frac{\Delta U}{k_BT}$. Number of realizations to obtain this quantity is typically $10^3$-$10^5$ in our simulations, depending on the barrier height.

As it is shown in Fig. \ref{fig3}(a), values of both $\tau_0$ and $\Delta U$ for the cis side are smaller than those of the trans side. In Fig. \ref{fig3}(a), product of $\tau_0$ and $\exp\left(\frac{\Delta U}{k_BT}\right)$ is also compared with $\tau$, which shows that they are of the same order of magnitude. In this figure, the size of error bars on $\tau$ and $\tau_0$ data points are of the same order of the data itself. Such large error bars are resulted from the exponential distribution of these quantities (see the inset of Fig. \ref{fig3}(b)). For any variable of exponential distribution, it is known that the mean and the standard deviation are of the same value. Actually, exponential distribution of entry time originates from the fact that consecutive polymer entries into the channel are independent events \cite{meller-branton}.

\section{Theory} \label{theory}
In this section, we present a simple scaling theory for the model system introduced in the previous section which will be tested with the simulation results in the next section. An important point in this theory is that the electric field compresses the polymer to the wall before its entry to the channel. Investigation of the polymer conformation in this compressed state is necessary for obtaining the free-energy barrier, $\Delta U$, and the polymer attempt frequency, $\tau_0^{-1}$.
\begin{figure}
\includegraphics[scale=0.3]{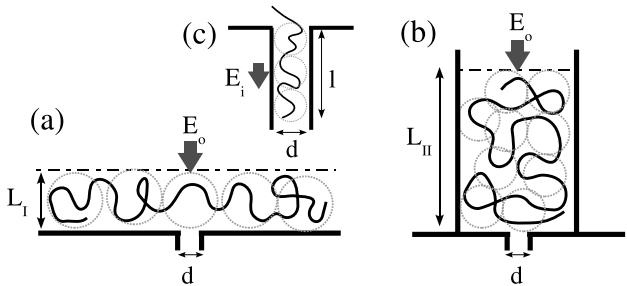}
\caption{\label{fig:epsart} Schematic of a polymer under the electric field of strength
$E_o$ behind a channel of diameter $d$ in a wall. (a) The polymer can be
viewed as a 2D chain of blobs of size $L_I$. (b) A polymer which is
also confined laterally by a cylinder of diameter $D$. The polymer can
be considered as a chain of blobs closely packed inside a cylinder
of height $L_{II}$. (c) A segment of the polymer entered to the channel
inside which the electric field is of strength $E_i$. } \label{fig4}
\end{figure}

As case $I$, consider a polymer with $N$ monomers of length $b$ and charge $q$, which is pushed to a wall by an applied electric field. The electric field tends to move the polymer toward the wall and decrease its electric energy. On the other hand, entropy resists the polymer confinement in a small volume near the wall. Competition between these two effects determines the polymer layer thickness, $L_I$ (see Fig. \ref{fig4} (a)). Actually, the electric field effect on the polymer is similar to the polymer adsorption to the wall, which is modeled with the polymer confinement between two walls \cite{rubinstein}. Here, one can introduce the blob size, a length scale in the system smaller than which the polymer has its unperturbed statistics and beyond it, the confinement by the wall dominates. So, the polymer can be modeled as a 2D chain of blobs of size $L_I$. Inside these blobs the Flory statistics reads $L_I\sim bg_I^{\nu}$ in which, $g_I$ is the number of monomers in each blob. Free-energy of the polymer under the electric field is of the order of $k_BT$ per blob; $F_{ent}\sim k_BT\frac{N}{g_I}\sim  k_BTN\left(\frac{b}{L_I}\right)^{\frac{1}{\nu}}$.

As case $II$, consider a similar polymer which is also confined from sides by a cylinder of diameter $D$ (see Fig. \ref{fig4} (b)) . In this case also, the effect of the electric field is equivalent to confining the polymer between two parallel walls. Because of the simultaneous confining effects of the electric field and the cylinder, this situation is similar to a polymer confined in a closed cavity. In this case, blobs can be defined that inside them the Flory statistic governs, $\xi\sim bg_{II}^{\nu}$. These blobs are closely packed inside the cavity, $\frac{N}{g}\xi^3\sim\Omega$. Here $\xi$ and $g_{II}$ are the blob size and the number of monomers in each blob, respectively, and $\Omega\sim L_{II}D^2$ is the cavity volume \cite{sakaue}. These two relations give the number of monomers in each blob $g_{II}\sim\left(\frac{\Omega}{Nb^3}\right)^{\frac{1}{3\nu-1}}$, and the polymer free-energy is obtained as $F_{ent}\sim k_BT\frac{N}{g_{II}}\sim k_BT\left(\frac{N^{3\nu}b^3}{\Omega}\right)^{\frac{1}{3\nu-1}}$.

We obtain the electric energy of the polymer layer, $F_{elc}$, in both cases with a mean field approximation. A uniform distribution is assumed for monomers in the layer of thickness $L_m$ $(m=I,II)$. Taking zero of the electric energy on the wall, we obtain $F_{elc}\sim NqE_oL_m$.

Because of the long entry time, polymer reaches to an equilibrium state before its entry to the channel. So we can find the thickness of the polymer layer in both cases by minimizing the sum of electric and entropic energies of the polymer, $F_{ent}+F_{elc}$, with respect to $L_m$:
\begin{equation}  \label{L}
L_I \sim b\left(\frac{qE_ob}{k_BT}\right)^{\frac{-\nu}{1+\nu}},  \\
L_{II} \sim b\left(\frac{qE_ob}{k_BT}\right)^{\frac{1-3\nu}{3\nu}}
\left(\frac{Nb^2}{D^2}\right)^{\frac{1}{3\nu}}.
\end{equation}

Free-energy per monomer before entry to the channel is $\sim k_BT/g_m$ and the number of monomers in each blob is obtained as
\begin{equation}
g_I \sim \left(\frac{qE_ob}{k_BT}\right)^{\frac{-1}{1+\nu}}, \\
g_{II} \sim
\left(\frac{qE_ob}{k_BT}\frac{Nb^2}{D^2}\right)^{\frac{-1}{3\nu}}.
\label{g}
\end{equation}

Consider a polymer in each one of cases $I$ and $II$ that enters a channel of diameter $d$ on the wall. Inside the channel, we take the polymer as a chain of blobs of diameter $d$ (see Fig. \ref{fig4} (c)). Statistics of the polymer segment inside a blob is unperturbed and $d\sim bg_{in}^{\nu}$ in which $g_{in}$ is the number of monomers in each blob inside the channel. Entropic energy of each monomer inside the channel is $\sim \frac{k_BT}{g_{in}}$, where $g_{in} \sim \left(\frac{d}{b}\right)^{\frac{1}{\nu}}$. A chain of $n$ monomers has a length $l\sim \frac{n}{g_{in}}d\sim n(\frac{b}{d})^{1/\nu}d$ inside the channel.

Hence, total change in the polymer free-energy after entry of $n$ monomers into the channel is $\Delta F_m \sim k_BT \left(\left(\frac{b}{d}\right)^{\frac{1}{\nu}}-\frac{1}{g_m}\right)n-n^2qE_i(\frac{b}{d})^{1/\nu}d$. In this relation, the term linear in $n$ comes from the entropic energy and the quadratic term is the electric energy. Maximum of $\Delta F_m$ with respect to $n$ gives the height of the free-energy barrier, $\Delta U_m$, which determines the rate of the polymer entry into the channel:
\begin{equation} \label{dF}
\frac{\Delta U_m}{k_BT}\sim
\left(\left(\frac{b}{d}\right)^{\frac{1}{\nu}}-\frac{1}{g_m}\right)^2
\frac{k_BT}{qE_id}\left(\frac{d}{b}\right)^{1/\nu}.
\end{equation}

The main dependence of $\Delta U_m$ on $E$ comes from the $E^{-1}$ term. The prefactor which contains $g$ has a weak dependence on the electric field and over small intervals of the field strength this term is almost constant. Such linear behavior of the energy barrier with $E^{-1}$ can be seen in the inset of Fig. \ref{fig3}(a). But over large electric field intervals, the changes coming from $g$-dependent prefactor is considerable and $\Delta U$ versus $E^{-1}$ would show deviations from the linear behavior. This has also been observed in the experiment \cite{craighead}.

In the inset of Fig. \ref{fig3}(a), the slope of the fitted line for case $I$ is larger than that of case $II$. This is because of the $g$-dependent prefactor in Eq. \ref{dF} and larger $g$ for case $I$.

Equation \ref{dF} can also be applied to free-energy barrier in entropic trapping experiments \cite{craighead}. In these experiments, the mobility of the charged polymer under electric field is measured in a channel with alternative wide and narrow regions. This mobility depends on the polymer trap time at the entrance of the narrow regions. Trap times are determined by the free-energy barrier against the entry of the polymer from the wide parts into the narrow parts of the channel. This method has been used for polymer separation \cite{separation}. It has been shown that in the experiment conditions, the free-energy barrier does not depend on the polymer length, $N$, and dependence of the trap time on $N$ comes only from the dependence of the polymer attempt time for passing the barrier on the polymer length. This time decreases with increasing the polymer length \cite{craighead,attempt-time,entropic}. The interesting point in Eq. \ref{dF} is dependence of the free-energy barrier on the polymer length in case $II$. In this case, $\Delta U$ decreases with increasing the polymer length, $N$. Thus by tuning the channel parameters the situation for the polymer before its entry to the channel narrow part could be similar to case $II$. In this way, dependence of the trap time on the polymer length would get stronger and may lead to more optimized polymer separation.

The number of monomers in each blob, $g_m$, can not be smaller than $1$ or larger than the total number of the monomers, $N$. In weaker fields (or shorter polymer lengths), the field is not strong enough to confine the polymer in a layer close to the wall and $g_m>N$. As a result, the polymer takes its unperturbed conformation and the radius of gyration is $R\sim bN^{\nu}$. In the other limit, $g<1$ means that the polymer has lost all of its entropy under the electric field and it enters the channel without feeling a free-energy barrier.
These two limits have also been observed in experiments. A crossover voltage below which no entry happens and another voltage limit above which the entry time changes linearly with the applied voltage have been reported \cite{bayley,pelta-auvray}.

\section{Simulation Results} \label{simulation}
\subsection{Statics of the polymer before entry to the channel}
In simulations for checking the theoretical results on the polymer statics before its entry to the channel in both cases $I$ and $II$, we block the channel entrance and let the polymer to equilibrate under the electric field. Then we measure the radius of gyration of the polymer parallel and perpendicular to the wall and find their scaling exponents with the polymer length and the electric field strength.

In case $I$, the polymer feels no lateral confinement on the wall. In this case, the blobs can be considered as the effective monomers of a 2D polymer. The blobs have a self-avoiding random walk on the wall and the radius of gyration of the polymer parallel to the wall is
\begin{equation} \label{RII}
R_{||}\sim L\left(\frac{N}{g_I}\right)^{\frac{3}{4}}\sim N^{\frac{3}{4}}b \left(\frac{b}{L}\right)^{\frac{1}{4}}.
\end{equation}
In this relation, $g_I$ is substituted from Eq. \ref{g} and $\frac{3}{4}$ is the Flory exponent in 2D. If the diameter of the confining cylinder is larger than this radius of gyration, $D>R_{\parallel}$, it has no confining effect on the polymer and the situation is similar to case $I$. But if the cross section of the cylinder is smaller than the monomers total area, $D^2<Nb^2$, the monomers pile on each other and case $II$ is relevant. Thus, for a given diameter of the confining cylinder, short and long polymers (or weak and strong fields) are equivalent to cases $I$ and $II$, respectively. One expects the crossover between these regimes to occur in the interval $\left(\frac{D}{b}\right)^{\frac{4}{3}}<N<\left(\frac{D}{b}\right)^2$, when $\frac{qEb}{k_BT} \sim 1$.

\begin{figure}
\includegraphics[scale=0.5]{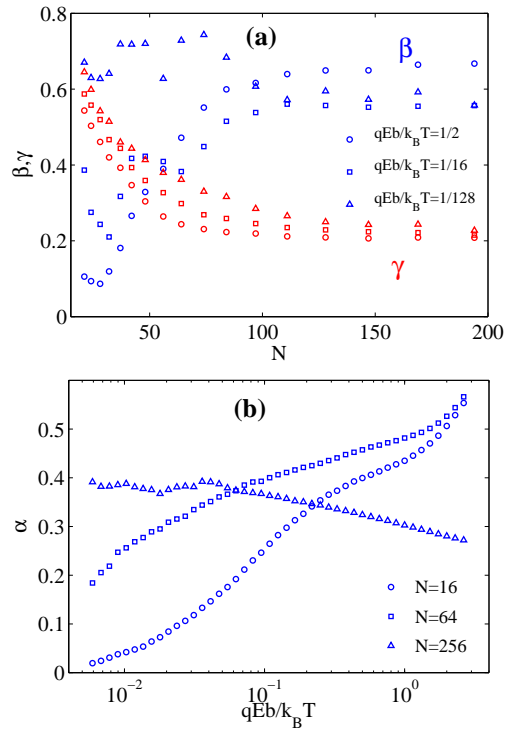}
\caption{ (Color online) (a) Exponents $\beta$ and $\gamma$ in relations $L\propto N^{\beta}$ and $R_{\parallel}L^{\frac{1}{4}}\propto N^{\gamma}$ for three different values of the field strengths. (b) Exponent $\alpha$ in  $L\propto E^{-\alpha}$ for three different polymer lengths.}
\label{fig5}
\end{figure}

According to Eqs. \ref{L} and \ref{RII}, one can write the relations $L\propto N^{\beta}$ and $R_{\parallel}L^{\frac{1}{4}}\propto N^{\gamma}$ for dependence of perpendicular and parallel components of the radius of gyration of the polymer on the number of its monomers. Then, it is obtained that $\beta_I=0$ and $\gamma_I=0.75$ for  case $I$, and $\beta_{II}=\frac{1}{3\nu}\simeq 0.6$ and $\gamma_{II}=\frac{1}{12\nu}\simeq 0.14$ for case $II$. These exponents which are calculated from the local slopes of the log-log plots of $L$ and $R_{\parallel}L^{\frac{1}{4}}$ versus $N$ are shown in Fig. \ref{fig5}(a). For the electric fields $\frac{qEb}{k_BT}=\frac{1}{2},\frac{1}{16}$ with increasing $N$, crossover from case $I$ to case $II$ can be seen. The weaker electric field $\frac{qEb}{k_BT}=\frac{1}{128}$ is out of the validity limits of the theory and we have $\gamma\sim \beta\sim \nu=0.6$ for all the polymer lengths. In these simulations, the diameter of the confining cylinder is $D=12\sigma$. With this diameter, shorter polymers do not feel the confining effect of the cylinder (case $I$). Longer polymers however, correspond to case $II$. The crossover between the two cases happens at $27<N<144$, in agreement with the theory.

Using Eq. \ref{L}, dependence of the radius of gyration of the polymer perpendicular to the wall on the electric field strength can be written as $L\propto E^{-\alpha}$. For cases $I$ and $II$, $\alpha_I=0.37$ and $\alpha_{II}=0.44$ are obtained.  This exponent is calculated from the local slope of the log-log plot of $L$ versus $E$ (see Fig. \ref{fig5}(b)). For the polymer length $N=256$ in the weaker fields, the exponent $\alpha$ is very close to the expected value $0.4$ (note  that values of the exponent $\alpha$ are very close to each other in these cases and distinguishing between them is not possible in the simulation). With increasing the electric field, the exponent $\alpha$ decreases, which is the result of approaching the validity limit of the theory, $g<1$. Assuming that the scaling constant in Eq. \ref{g} is of the order 1, $g_{II}=\left(\frac{qE_ob}{k_BT}\frac{Nb^2}{D^2}\right)^{\frac{-1}{3\nu}}=1$ gives the value $\frac{qEb}{k_BT}=0.5$ for this limit. For the two shorter polymers, it is seen that in the range of $0.03<\frac{qEb}{k_BT}<1$ for $N=64$ and in the range of $0.3<\frac{qEb}{k_BT}<1$ for $N=16$, the exponent $\alpha$ is between $0.3$ and $0.5$ and the curve slope is smaller. But despite the theory, no plateau can be seen in this curve, which may be a result of the failure of the mean field approximation.

\begin{figure}
\includegraphics[scale=0.5]{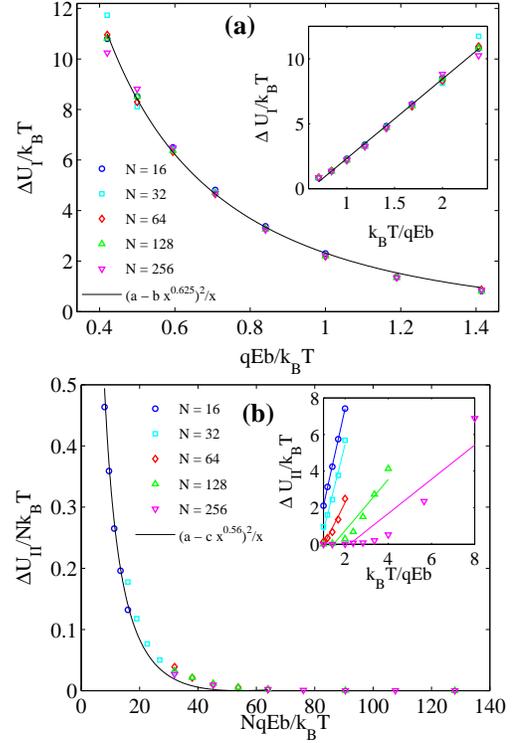}
\caption{(Color online) (a) , (b) Symbols are simulation results for free-energy barrier versus the electric field for different values of the polymer length in cases $I$ and $II$. In (b) the axes are scaled such that all the data sets fall on a master curve. The solid line is the fitted function of the theory, Eq. \ref{dF}, to the simulation data. Insets show free-energy barrier versus $E^{-1}$. Deviations from linear behavior can be seen over large intervals of the electric field, specially in case $II$.}
\label{fig6}
\end{figure}

\subsection{Free-energy barrier}
To check the main result of our scaling theory, Eq. \ref{dF}, we calculate the height of the free-energy barrier with the same method as described in Sec. \ref{model} for different values of the polymer length and the electric field strength (see Fig. \ref{fig6}). The diameter and the length of the channel in these simulations are  $d=1.4\sigma$ and $H=6\sigma$ and the
diameter of the confining cylinder is $D=8\sigma$. As it can be seen in this figure, our scaling theory can explain the exponents but not the coefficients. We use the function $(a-bx^{0.6})^2/x$ for fitting to $\frac{\Delta U}{k_BT}$ versus $\frac{qEb}{k_BT}$ for case $I$ and the function $(a-cx^{0.6})^2/x$ for fitting to $\frac{\Delta U}{Nk_BT}$ versus $\frac{NqEb}{k_BT}$ for case $II$. $a$, $b$ and $c$ are fit parameters. The parameter $a$ depends on the ratio of the monomer size to the diameter of the channel. It is taken the same in separate fittings to the data sets of cases $I$ and $II$. As it can be seen, there is a good agreement between simulation and theory and the fit parameters are close to their expected values. Taking $d\sim b$, the expected values are $a\sim \left(\frac{b}{d}\right)^{\frac{1}{2\nu}+\frac{1}{2}} \sim 1$, $b\sim \left(\frac{b}{d}\right)^{-\frac{1}{2\nu}+\frac{1}{2}} \sim 1$ and $c\sim \left(\frac{b}{D}\right)^{\frac{2}{3\nu}} \left(\frac{b}{d}\right)^{-\frac{1}{2\nu}+\frac{1}{2}} \sim 0.1$ in reasonably well agreement with the values obtained from fitting to the simulation data, $a=3.02$, $b=1.50$ and $c=0.32$.

In the insets of Fig. \ref{fig6} (a) and (b), deviation of $\Delta U$ versus $E^{-1}$ from linear form in large intervals of $E^{-1}$, no dependence of the free-energy barrier on the polymer length in case $I$ and its decrease with the polymer length in case $II$ can be seen.

\subsection{Polymer attempt time for crossing the barrier, $\tau_0$}
\begin{figure}
\includegraphics[scale=0.51]{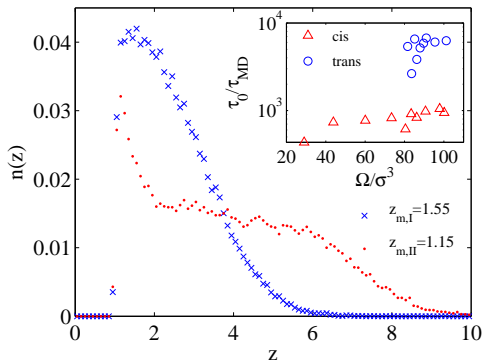}
\caption{(Color online) Distribution of $z$ component for one of end monomers of the polymer for cases $I$ and $II$. The values of $z$ components corresponding to the maximum of the distributions are $z_{m,I}=1.55$ and $z_{m,II}=1.15$. Inset: $\tau_0$ versus $\Omega=LD^2$. Although the values of $\Omega$ for cases $I$ and $II$ are close to each other, the values of $\tau_0$ for the two cases are very different. It may be resulted from the failure of the mean-field assumption for the distribution of the monomers in our theory.}
\label{fig7}
\end{figure}

According to the Rouse model, internal monomers of the polymer have an anomalous diffusion and their mean square displacement changes with time as $\left<\Delta r^2\right>\propto \Delta t^{0.5}$ \cite{rubinstein}. In our model, one can consider the end monomer of the polymer as a particle that is anomalously diffusing in a closed cavity of volume $\Omega=LD^2$, with $L$ the radius of gyration of the polymer perpendicular to the wall. The mean polymer attempt time for passing the barrier is then equal to the mean first passage time for this particle. This time changes with the 4th power of the system characteristic length, $\Omega^{\frac{1}{3}}$: $\tau_0\propto\Omega^{\frac{4}{3}}$ \cite{gitterman}.
Our simulation results for the attempt time are very limited and scattered and we cannot reject or prove this relation.
But a point which can be emphasized here is the large difference between the polymer attempt times from the cis and trans sides, despite the similar values of $\Omega$ for the two sides (see the inset of Fig. \ref{fig7}). Actually, in the trans side, the end monomers do not get close to the wall easily for entropic reasons. But in the cis side, the polymer is confined in a closed volume and the pressure of the other monomers push the end monomers toward the wall as is shown in Fig. \ref{fig7}. In this figure, there is a pick in distribution of the position of the end monomer on the wall in case $II$, which is not seen for case $I$. This point can explain the large difference between the two cases and shows the failure of the mean field approximation for this part of the problem.

\section{Comparison with experiment}  \label{experiment}
\subsection{Partition coefficient}
In the experiments, the partition coefficient is measured to calculate the confinement free-energy of the polymer. The partition coefficient is the ratio of polymer concentrations inside the channel and in the bulk solution. A voltage difference is applied to keep the polymer in the channel and ion current through the channel is measured \cite{partitioning2,partitioning3}. Also, quantities such as polymer reaction rate with a residue inside the protein channel can be measured \cite{partitioning1}.

From our theory, we expect the confinement free-energy of a polymer of length $N$ inside a narrow channel of diameter $D$ under the electric field $E$ to obey the relation $\frac{\Delta F}{k_BT}\sim\frac{N}{g_{II}}\propto N^{1.6}E^{0.6}\left(\frac{b}{D}\right)^{1.1}$. This relation is valid for high salt concentrations (no long range electrostatic interaction between the monomers) and for polymer lengths and electric field strengths that are in the validity range of the theory.

There are numerous suggestions for dependence of the confinement free-energy on the polymer length in the literature. In Ref. \cite{partitioning2} exponents $1.6$, $1.35$ and $3.2$ are obtained for confinement inside different channels. Exponent $3.2$ is explained in Ref. \cite{sakaue} and it is attributed to the almost closed shape of the channel used in the experiment. The two other exponents are in good agreement with our theory. In Ref. \cite{partitioning1} it has been shown that the partition coefficient obeys the de Gennes theory. In this theory, the confinement free-energy of the polymer inside a cylinder without any electric field is $\frac{\Delta F}{k_BT}\sim N\left(\frac{b}{D}\right)^{\frac{5}{3}}$. The linear dependence of the free-energy on the polymer length in this experiment may be because of the low salt concentration, lower applied electric field or smaller polymer length that are beyond the polymer compressing limit by the electric field and hence beyond the validity range of our theory.
\begin{figure}
\includegraphics[scale=0.5]{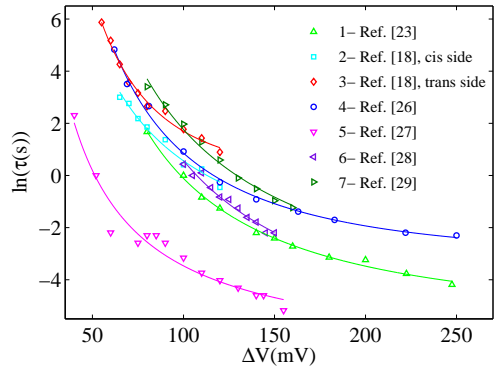}
\caption{(Color online) Logarithm of capture time versus $\Delta V$
extracted from referenced articles (symbols) and fitted equation of
our theory, Eq. \ref{dF} (solid lines). All the data sets are from polymer
translocation experiments through the $\alpha$-hemolysin channel. As it can
be seen, all data sets are followed well by the suggested function. Fit
parameters are shown in Table \ref{table2}.}
\label{fig8}
\end{figure}

\subsection{Dependence of the entry time on the applied voltage} \label{exp-entry}
According to Eq. \ref{arrhenius}, the entry time depends on the free-energy barrier height as well as the polymer attempt time for passing the barrier.
Considering the weak dependence of $\tau_0$ on the applied voltage (see Fig. \ref{fig3}(a)), dependence of the entry time on the applied electric field for both cases $I$ and $II$ can be written using Eqs. \ref{arrhenius} and \ref{dF} as
\begin{equation} \label{comp3}
\ln \tau =\zeta + \left( \eta -\lambda
\left(\frac{qE_ib}{k_BT}\right)^{0.6}\right)^2\frac{k_BT}{qE_ib}.
\end{equation}
$\eta$, $\lambda$ and $\zeta$ are parameters independent of the applied voltage. Most of the applied voltage drops along the channel and dependence of the electric field inside the channel on the voltage reads $E_i=\frac{\Delta V}{H}$, where $H$ is the channel length. Here, it is assumed that the electric field outside the channel is uniform and has a linear relation with the field strength inside the channel, $E_o=\delta E_i$, where $\delta$ is a constant parameter. We obtain $b$ and $q$ in Eq. \ref{comp3} from coarse-graining rule of the polymer. We take the coarse-grained monomers size, $b$, the same as the polymer diameter in  the experiment. The charge of coarse-grained monomers, $q$, is determined from known linear charge density of the polymer. Parameters related to several flexible polymers and their coarse-grained monomers are shown in Table \ref{table1}.

\begin{table}
\caption{\label{table1} Parameters of the three polymers used in the experiments and the corresponding course-grained parameters. The polymers are single-stranded DNA, dextran sulfate and poly(styrene) sulfonate.}
\begin{ruledtabular}
\begin{tabular}{lccc}
&ssDNA  &DS &PSS \\
\hline
real monomers separation (\AA) &3.3 &2.5 &2.5 \\
real monomers charge (e) &1 &2 &1 \\
polymer diameter (\AA) &10 &7 &8 \\
coarse-grained monomer size (b) (\AA) &10 &7 &8\\
coarse-grained monomer charge (q) (e) &3 &5.6 &3.2  \\
polymer persistence length (\AA) &15 &16 &14   \\
\end{tabular}
\end{ruledtabular}
\end{table}

According to Eqs. \ref{arrhenius} and \ref{dF}, $\zeta \sim \ln \tau_0$, $\eta\sim \left( \frac{d}{b}
\right)^{\frac{1}{2}+\frac{1}{2\nu}}$ for both cases $I$ and $II$, $\lambda\sim \left( \frac{d}{b} \right)^{\frac{1}{2}-\frac{1}{2\nu}}\delta^{0.6}$ for case $I$ and $\lambda\sim \left( \frac{d}{b}\right)^{\frac{1}{2}-\frac{1}{2\nu}}\delta^{0.6} \left(\frac{Nb^2}{D^2} \right)^{0.6}$ for case $II$. We use these quantities as fit parameters to the experimental results as shown in Fig. \ref{fig8}. All these results are related to polymer passage through the $\alpha$-hemolysin channel, so $H=10nm$. Fit parameters are shown in Table \ref{table2}.

Because of the complex shape of the $\alpha$-hemolysin channel, the profile of the electric field strength in the channel vicinity is not very clear in these experiments and we can only investigate the fit parameters qualitatively. $\zeta$ depends on $\tau_0$ and hence on the polymer concentration in the solution and the polymer length. $\eta$ depends on the ratio of diameters of the polymer and the channel. In addition, $\eta$ could be dependent on the electrostatic interactions between the polymer and the channel. Regarding almost the same charge and diameter of the polymers used in the experiments listed in Table \ref{table2}, and also considering the same protein channel used in these experiments, the same value of $\eta$ for all data sets is acceptable. $\lambda$ is the main parameter of interest here. It is a measure of polymer compression before its entry to the channel. Diameter of the $\alpha$-hemolysin channel in the trans side is smaller than the cis side. This causes a higher resistance for this part of the channel and a higher voltage drop and stronger electric field inside it. The stronger electric field leads to stronger compression of the polymer. It can be seen that $\lambda$ is larger for the data sets which are for translocation from the trans side compared to those for the cis side. In addition, in translocation from the cis side, when the applied voltage is higher, compression effect is stronger and the value of $\lambda$ is larger. The data sets 1 and 4 are for high voltages and data sets 2, 5 and 7 are for low voltages (all of them are for the cis side).
The point which should be noted here is that according to Ref. \cite{bayley}, in translocation from the cis side, the polymer is not compressed inside the wide part of the channel. So, entries from both cis and trans sides may correspond to case $I$ of our theory, not case $II$.
\begin{table}
\caption{\label{table2}Parameters $\zeta$, $\eta$, and $\lambda$
from fitting of Eq. \ref{comp3} to experimental data. Result is shown in Fig. \ref{fig8}.}
\begin{ruledtabular}
\begin{tabular}{ccccc}
&Experimental data  &$\zeta$ & $\eta$
&$\lambda$ \\
\hline
1 &ssDNA, cis side (Ref. \cite{bayley})& -5.42 & 3.14 & 0.60 \\
2 &ssDNA, cis side (Ref. \cite{henrickson})& -4.33 & 2.38 & 0 \\
3 &ssDNA, trans side (Ref. \cite{henrickson})& 0.27 & 3.22 & 1.77 \\
4 &ssDNA, cis side (Ref. \cite{meller-branton}) & -3.01 & 3.11 & 0.90\\
5 &DS, cis side (Ref. \cite{pelta-auvray})& -7.09 & 2.35 & 0.01  \\
6 &DS, trans side (Ref. \cite{trans-cis})& -6.13 & 3.85 & 0.51  \\
7 &PSS, cis side (Ref. \cite{chen})& -6.01 & 2.83 & 0  \\
\end{tabular}
\end{ruledtabular}
\end{table}

\section{Summary and Discussion} \label{summary}
Entry of a flexible polymer into a narrow channel which is the most time-consuming stage of the electric-field-driven polymer translocation through nanoscale channels in the barrier-limited regime has been studied theoretically and using MD simulations. Two different geometries for the channel entrance have been considered to investigate the effect of the lateral confinement of the polymer before its entry to the channel narrow part. The height of the free-energy barrier has been obtained as a function of the strength of applied electric field and the length of the polymer for the two entrance geometries. There is a very good agreement between our theoretical and simulation results. Our suggestion for dependence of the polymer entry time on the applied voltage across the channel fits with the related experimental data and the behavior of the fit parameters are described. Our theory also explains the polymer partitioning experiments.

With the aim of describing the reported difference between polymer entry times from cis and trans sides of the asymmetric $\alpha$-hemolysin channel, MD simulation of polymer entry into a channel of dimensions very close to those of $\alpha$-hemolysin has been performed. The order of magnitude and the ratio of polymer entry times from cis and trans sides obtained from our simulations are in agreement with the corresponding experiment. According to our results, existence of a confining space prior to the channel narrow part in the cis side, very similar to the geometry of the $\alpha$-hemolysin channel, causes the same difference between the entry times from the cis and the trans sides as has been observed in the experiments.  
Free-energy landscape seen by the polymer when it enters from the cis side is shown schematically in Fig. \ref{fig1}. When the electric field is strong enough, after passing one step of the energy barrier (entering into the wide part), the polymer is trapped in a region where its free-energy is lower relative to the outside of the channel (the gray region in the lower panel of Fig. \ref{fig1}). It causes the passage from the two steps of the energy barrier to happen as two separate stages and the passage probability from the whole barrier not to be the product of passage probabilities from the two individual steps. If the diameter and/or the length of the wide part or the electric field strength is small enough, the region in which the polymer has a lower free-energy disappears. Then the total height of the free-energy barrier would determine the polymer entry time. We have observed such behaviors in our test simulations.

The channel projection out of the membrane has been shown that has not an important role in the polymer entry process from the cis side in our simulations. In our test simulations with the real structure of the $\alpha$-hemolysin channel taken from the protein data bank \cite{pdb}, we observed that the projected surface of the channel in the cis side has a larger diameter than the radius of gyration of the polymer (for a polymer of length $N=25$). So the projected part of the channel effectively interacts as a wall with the polymer. Therefore, in all our simulations we have put the wall at the channel ends in both the cis and trans sides. In addition, in Ref. \cite{ambjornsson} that the channel projection is suggested as a reason for difference between entry times from the cis and trans sides, the electric field outside the channel is taken to be zero. In our simulations however, electric field outside the channel compresses the polymer and the entropic cost of the attachment of the polymer end to the wall is small. The effect of the projected part may be important at weaker electric fields or for longer polymers.

To make the model used in our theory and simulation more similar to the real system in the experiment, two points should be considered. First, the electric field outside the channel should be calculated from more accurate theories \cite{grosberg} and also the field gradient inside the $\alpha$-hemolysin channel should be taken into account \cite{bayley2}. Second, some percentage of the polymers that enter from the cis side of the $\alpha$-hemolysin channel are compressed inside the wide part of the channel before their entry to the narrow part \cite{bayley}. In this situation, a portion of a long polymer would remain out of the wide part of the channel. This case is not studied in our theory. In this situation, the polymer is divided into two parts: one is compressed inside the cylinder and the other is free on the wall. The part outside the channel may act like a spring which drags the inner part out of the channel \cite{bezrukov}.

In many of the polymer translocation experiments, double stranded DNA which its bending rigidity is considerably larger than ssDNA has been used. A more complete theory is needed to be developed for semiflexible polymers. More complex regimes have been predicted for semiflexible polymers confined in a nanoscale space \cite{odijk}.

\begin{acknowledgements}
We would like to acknowledge two anonymous referees for their valuable comments and suggestions.
\end{acknowledgements}

\end{document}